\documentclass{article}
\usepackage{graphicx}
\usepackage{amsmath,amssymb}
\usepackage{graphics,color,array,calc,rotating,epsfig,psfrag}
\numberwithin{equation}{section}
\usepackage{subcaption}
\usepackage{cite}
\usepackage{bm}
\usepackage{dcolumn}
\usepackage{float}
\usepackage{amsfonts}
\usepackage[mathscr]{eucal}
\def\be{\begin{equation}} \def\ee{\end{equation}}
\def\bea{\begin{eqnarray}} \def\eea{\end{eqnarray}}

\newcommand\prt{\partial}

\begin{document}
\baselineskip 18pt%
\begin{titlepage}
\vspace*{1mm}%
\hfill%
\vspace*{15mm}%
\hfill
\vbox{
    \halign{#\hfil         \cr
%           hep-th/yymmnnn\cr
%         IPM/P-2009/025  \cr
          } % end of \halign
      }  % end of \vbox
\vspace*{20mm}

\begin{center}
{\large {\bf  Joule-Thomson expansion of charged AdS black holes in
Rainbow gravity}}\\
\vspace*{5mm}
{Davood Mahdavian Yekta$^{*}$\footnote{d.mahdavian@hsu.ac.ir}, Arezoo Hadikhani$^{*}$, \"{O}zg\"{u}r \"{O}kc\"{u}$^{\dagger}$\footnote{ozgur.okcu@ogr.iu.edu.tr}}\\
\vspace*{0.2cm}
{$^{*}$ \em{Department of Physics, Hakim Sabzevari University, Sabzevar, P.O.Box 397, Iran}}\\
{$^{\dagger}$ \em{Department of Physics, Faculty of Science, Istanbul University, Istanbul, 34134, Turkey}}\\
\vspace*{1cm}
\end{center}

\begin{abstract}
In this letter we investigate the throttling process of the charged Anti-de Sitter (AdS) black holes in the rainbow gravity. In the extended phase space of these black holes, the cosmological constant plays the role of a varying thermodynamic pressure and the black hole mass is identified with the thermodynamic enthalpy. We derive exact expressions for the Joule-Thomson coefficient and the inversion temperature in terms of black hole parameters and constants of rainbow gravity analytically, and then perform a numerical analysis for the isenthalpic and inversion curves of charged AdS black holes. Our calculations show all quantities are sensitive to rainbow parameter $\eta$. We also discuss the isenthalpic curves for different values of the black hole mass.

\end{abstract}
\end{titlepage}

%%%%%%%%%%%%%%%%%%%%%%%%%%%%%%%%%%%
\section{Introduction}
There are two significances in General Relativity (GR) which cause it not to be a fundamental theory of gravity and needs to be modified or extended. The idea of short range or Ultra Violet (UV) modification of GR, which arises from the non-renormalizability of the theory \cite{Stelle:1976gc}, and the other is large scale or Infra Red (IR) modification of GR, which might be the explanation of the observed late-time universe acceleration\cite{Capozziello:2011et} or of the inflationary stage \cite{Nojiri:2003ft}. A large amount of research has been devoted to both implications but here we focus on the UV modification. Though the addition of higher curvature terms to the Einstein-Hilbert action establishes renormalizability, but this leads to the ghosts problem, that is the equations of motion involve higher-order time derivatives\cite{Stelle:1976gc,Stelle:1977ry}. Another approach including higher spatial derivatives only, and motivated by the Lifshitz theory was constructed by Horava \cite{Horava:2008ih,Horava:2009uw}. The Horava-Lifshitz gravity includes different anisotropic scaling for the space and time and is regarded as a UV completion of GR. \footnote{Horava-Lifshitz gravity is known to suffer from perturbative instability in the IR, ultimately due to an extra mode which comes from the explicit breaking of general covariance (because of higher spatial derivatives but not higher time derivatives). One of the ways to fix this is to have Horava gravity emerging dynamically in the UV while preserving Lorentz invariance (or rather, not having a preferred foliation) in the IR \cite{Cognola:2016gjy}.}

However, there is an alternative UV completion of GR in which instead of modifying the action, the spacetime metric is deformed. This construction named Rainbow  Gravity (RG) \cite{Magueijo:2001cr}. This deformation exhibits a different treatment between space and time in the UV limit, nearly the Planck scale, depending on the energy of the particle probing the spacetime, while at low energy limit one recovers the standard form of the metric in GR. This deformation has been shown to cure divergences avoiding any renormalization scheme \cite{Garattini:2011kp,Garattini:2012ec}. As the standard energy-momentum dispersion relation depends on the Lorentz symmetry, it is expected that the standard energy-momentum dispersion relation will also get modified in the ultraviolet limit. The modification of the standard energy-momentum dispersion relation has motivated the development of doubly special relativity (DSR)\cite{AmelinoCamelia:1996pj,AmelinoCamelia:1997gz,AmelinoCamelia:2000ge,Bruno:2001mw}. In fact, DSR is an extension of special relativity which has two upper bounds; the velocity of light and Planck energy.
The generalization of this theory to general relativity is RG \cite{Magueijo:2002am,Magueijo:2002xx,Magueijo:2004vv}.

On the other hand, thermodynamics of black holes \cite{Bekenstein:1972tm,Bardeen:1973gs} has fundamental connections among the classical thermodynamics, general relativity, and quantum mechanics. More specifically, due to the development of AdS/CFT conjecture\cite{Maldacena:1997re,Gubser:1998bc}, this connection has been deepened and a lot of attention has been attracted to AdS black holes. The relation between geometrical properties of the event horizon and thermodynamic quantities provides a clear indication that there is a relation between properties of the spacetime geometry and some kind of quantum physics\cite{Hawking:1974rv,Hawking:1974sw,Hawking:1982dh}.
Chamblin and et al. have shown in Refs. \cite{Chamblin:1999tk,Chamblin:1999hg} that charged AdS black holes have rich phase structures.  This phase transition is analogous to a van der Waals liquid-gas phase transition \cite{Banerjee:2011raa} where the cosmological constant is identified as pressure \cite{Kastor:2009wy}. Kubiznak and et al. studied the P-V critical behavior and critical exponents in Refs. \cite{Kubiznak:2012wp,Hansen:2016ayo,Kubiznak:2016qmn} that corroborates deep analogy between a charged AdS black hole and a van der Waals fluid\footnote{Studies about the reentrant phase transition and triple point in the extended phase space have been done in Refs. \cite{Altamirano:2013uqa,Frassino:2014pha,Wei:2014hba,Hennigar:2015wxa}.}, such that a lot of attention has been turned into the study of extended phase space thermodynamics and P-V criticality of black holes in different theories such as Born-Infeld theory \cite{Gunasekaran:2012dq}, Non-linear Maxwell theory \cite{Hendi:2012um,Dehyadegari:2016nkd}, Gauss-Bonnet theory \cite{Cai:2013qga}, and higher order gravities \cite{Sherkatghanad:2014hda,Hendi:2018xuy}.

Another process comes from this analogy is the Joule-Thomson (JT) expansion. In JT expansion a gas at a high pressure passes through a valve or porous plug to a low pressure section such that during the process enthalpy is unchanged and the process is an adiabatic expansion. The interest of study this process for black holes returns to small rate of black hole Hawking radiation which can be regarded as an adiabatic expansion though there is no porous plug. Recently, the JT expansion process in the case of charged AdS black holes has been studied in Ref. \cite{Okcu:2016tgt} and a few later the authors also considered this effect for the Kerr-AdS black holes \cite{Okcu:2017qgo}. The computations for Kerr-Newman AdS black holes have been done in Ref. \cite{Zhao:2018kpz} and the generalizations to charged AdS solutions in the quintessence  and monopole black holes have been done in Refs. \cite{Ghaffarnejad:2018exz,AhmedRizwan:2019yxk}. Higher dimensional charged AdS black holes have been argued in Ref. \cite{Mo:2018rgq}. JT effect has been also considered in modified theories such as Lovelock gravity \cite{Mo:2018qkt}, $f(R)$ gravity \cite{Chabab:2018zix}, Gauss-Bonnet gravity \cite{Lan:2018nnp}, nonlinear electrodynamics\cite{Kuang:2018goo}, Einstein-Maxwell-Axion and massive gravity\cite{Cisterna:2018jqg}, and Bardeen theory \cite{Li:2019jcd}.

In this letter, we are curious to study the JT expansion for charged AdS black holes in RG. This generalization is of physical significance since varieties of
intriguing thermodynamic properties have been disclosed for this solution in RG  \cite{Ling:2005bp,Galan:2006by,Ali:2014yea,Gim:2015zra,Kim:2016qtp,Feng:2017gms,Dehghani:2018svw,Hendi:2018sbe,EslamPanah:2018ums}. It is naturally expected that this research may give rise to novel findings concerning the JT expansion. The main motivation for studying black holes in RG is to consider the quantum corrections on the classical perspectives. The
idea of energy dependent spacetime is one of those quantum corrections. Considering
this energy dependent spacetime, we can venture on the effects it brings about on the
JT expansion of charged AdS black hole. Due to modification of the original surface gravity in RG, the rainbow Hawking temperature is very sensitive to the concrete expression of rainbow functions and consequently the entropy from the first law. We study the inversion temperature by using the equation of state in which the cooling-heating transition occurs in throttling process. In this treatment, the mass of black hole plays the role of enthalpy in the extended phase space \cite{Kastor:2009wy,Kubiznak:2012wp,Kubiznak:2016qmn,Hansen:2016ayo,Dolan:2011xt,Caceres:2016xjz}. We plot the inversion and isenthalpic curves for different values of parameters and discuss some of the relevant issues in RG.

%%%%%%%%%%%%%%%%%%%%%%%%%%%%%%%%%%%
%@@@@@@@@@@@@@@@@@@@@@@@@@@@@@@@@@@@@
\section{Charged Black Holes in Rainbow Gravity}
Lorentz symmetry is one of the most important symmetries in nature which might be violated in the UV limit\cite{Adams:2006sv,Gripaios:2004ms,Iengo:2009ix}. Since the standard energy-momentum dispersion relation depends on the Lorentz symmetry, it is expected that it will also get modified in the ultraviolet limit named doubly special relativity(DSR) \cite{AmelinoCamelia:1996pj,AmelinoCamelia:1997gz,AmelinoCamelia:2000ge,Bruno:2001mw}. On the other hand, RG is a generalization of DSR applied to curved spacetime \cite{Magueijo:2002am,Magueijo:2002xx,Magueijo:2004vv}. In DSR theory there is a maximum energy scale in the nature which is the Planck energy in accordance to modification in UV limit. The energy-momentum dispersion relation for a test particle of mass $m$ is given by
\be\label{dr} E^2\, \mathcal{F}^2 \left(E/E_{p},\eta\right)-p^2 \,\mathcal{G}^2 \left(E/E_{p},\eta \right) = m^2,\ee
where the two energy dependent rainbow functions satisfy in
\be\label{rbf1} \lim_{E/E_{p}\rightarrow 0} \mathcal{F}\left(E/E_{p},\eta\right)=1, \qquad  \lim_{E/E_{p}\rightarrow 0} \mathcal{G}\left(E/E_{p},\eta\right)=1\,, \ee
and $\eta$ is a dimensionless constant which we called it rainbow parameter. The relation (\ref{dr}) goes to the standard form when the energy of the test particle is much lower than the Planck scale. The rainbow metrics lead to a one parameter family of connections and curvature tensors such that Einstein’s equation becomes \cite{Ling:2005bp,Peng:2007nj}
\be\label{Eeq} G_{\mu\nu}\left(E/E_{p} \right)+\Lambda \left(E/E_{p} \right)\,g_{\mu\nu} \left(E/E_{p} \right)=8\pi G\left(E/E_{p} \right) T_{\mu\nu}\left(E/E_{p} \right),\ee
where $G_{\mu\nu}(E/E_{P})$ and $T_{\mu\nu}(E/E_{P})$ are energy-dependent Einstein and energy-momentum tensors, and $\Lambda(E/E_{P})$ and $G(E/E_{P})$ are energy-dependent cosmological and gravitational constants. In this work, we choose the RG functions discussed in Refs. \cite{AmelinoCamelia:1997gz,AmelinoCamelia:1996pj,Jacob:2010vr,Ali:2014qra,Hendi:2015hja} which are phenomenologically important,
\be\label{rbf2} \mathcal{F}\left(E/E_{p},\eta\right)=1, \qquad \mathcal{G}\left(E/E_{p},\eta\right)=\sqrt{1-\eta\,\left(E/E_{p}\right)^n }\,.\ee

The modified charged AdS black hole in RG is described by the following line element analogous to Schwarzschild black hole in Ref. \cite{Magueijo:2002xx}
\be\label{rgm} ds^2=-\frac{f(r)}{ \mathcal{F}^2} dt^2+\frac{1}{ f(r) \mathcal{G}^2} dr^2+\frac{r^2}{\mathcal{G}^2} d\Omega^2\,,\ee
where
\be f(r)=1-\frac{2 G M}{r}+\frac{Q^2}{r^2}+\frac{r^2}{l^2}\,.\ee
The parameters $M$ and $Q$ are the mass and charge of the black hole which to avoid the singularity we should have $M\ge Q$, and $l$ is the radius of AdS space related to the cosmological constant as $\Lambda=-\frac{3}{l^2}$. The location of event horizon is given by the largest real root of $f(r)=0$.

%@@@@@@@@@@@@@@@@@@@@@
\section{Thermodynamics of Charged AdS black hole}
  The study of thermodynamic properties of asymptotically AdS black holes dates back to the seminal work of Hawking and Page\cite{Hawking:1982dh} about the phase transition in Schwarzschild-AdS black holes and for more complicated backgrounds in \cite{Cvetic:1999ne,Cvetic:1999rb}. The thermodynamics of charged black holes which is our interest in this paper, have been considered extensively in Refs. \cite{Chamblin:1999tk,Chamblin:1999hg}. In particular, in the case of an asymptotically AdS black hole in four dimensions the cosmological constant $\Lambda$ is identified with the pressure by \cite{Kastor:2009wy}
  \be\label{pl} P=-\frac{\Lambda(0)}{8\pi}=\frac{3}{8\pi l^2}\,,\ee
and its conjugate variable in black hole thermodynamics is the volume
\be\label{vol} V=\left(\frac{\prt M}{\prt P}\right)_{S,Q}=\frac43 \pi r_{+}^3\,,\ee
where $r_{+}$ is the event horizon of black hole. The temperature of the modified black hole in (\ref{rgm}) can be calculated from the surface gravity $\kappa$ on the horizon \cite{Ling:2005bp}, namely
 \be\label{temp1} T=\frac{\kappa}{2\pi}=-\frac{1}{4\pi} \lim_{r\rightarrow r_{+}} \sqrt{\frac{-g^{11}}{g^{00}}} \,\frac{(g^{00})'}{g^{00}}\,, \ee
 where prime is the derivative with respect to the $r$.  According to the uncertainty principle, $\Delta p\ge \frac{1}{\Delta x}$ can be translated to a lower bound on the energy of a test particle \cite{Adler:2001vs,AmelinoCamelia:2004xx}
 \be\label{up} E\ge \frac{1}{\Delta x}\sim \frac{1}{r_{+}}.\ee

Without lose of generality and for later convenience we take $G=1$ and $n=2$ in (\ref{rbf2}), so by substituting dispersion relation (\ref{dr}) and uncertainty relation (\ref{up}) in (\ref{rbf2})  we obtain the rainbow function as
\be\label{rbf3} \mathcal{G}=\sqrt{1-\eta G_{p} m^2}\sqrt{\frac{r_{+}^2}{r_{+}^2+\eta G_{p}}}=\frac{1}{k} \sqrt{\frac{r_{+}^2}{r_{+}^2+\eta G_{p}}}\,,\ee
where $G_{p}=1/E_{p}^2$, $k=[1-\eta G_{p} m^2]^{-1/2}$, and $m$ is the mass of test particle. Thus the temperature from (\ref{temp1}) is given by
\be\label{temp2} T=\frac{1}{4\pi k \sqrt{r_{+}^2+\eta G_{p}}}\,\frac{(8\pi P\,r_{+}^4+r_{+}^2-Q^2)}{r_{+}^2}\,.\ee
The black hole's mass can be easily calculated from the condition $f(r)=0$ in terms of horizon radius
\be\label{mass} M=\frac{3Q^2+3r_{+}^2+8\pi P r_{+}^4}{6r_{+}}\,,\ee
in which we have used the relation (\ref{pl}). In the previous section we assert that the area of horizon will be corrected when we use the modified metric (\ref{rgm}) \cite{Kim:2016qtp}. So, the modified entropy is as follows
\be\label{ent} S=\pi k\, r_{+} \sqrt{r_{+}^2+\eta G_{p}}+\pi k \,\eta \,G_{p} \ln{(\sqrt{r_{+}^2+\eta G_{p}}+r_{+})}\,,\ee
and we have checked that the quantities given in (\ref{temp2})-(\ref{ent}) satisfy the first law of black hole thermodynamic
\be\label{FL} T=\left(\frac{\prt M}{\prt S}\right)_{Q,P},\ee
for constant $Q$ and $P$. The pressure of charged AdS black holes in RG has been calculated in \cite{Li:2018gwf}
\be\label{press} P=\frac{k}{2}\,\sqrt{\frac{r_{+}^2+\eta G_{p}}{r_{+}^4}}\,T+\frac{Q^2-r_{+}^2}{8\pi r_{+}^4}\,,\ee
where in the limit $\eta=0$ it reduces to (3.10) in Ref. \cite{Kubiznak:2012wp}. We can also rewrite the relation (\ref{press}) as the equation of state by substituting $r_{+}$ in terms of volume $V$ from (\ref{vol}).
%@@@@@@@@@@@@@@@@@@
\section{Joule-Thomson Expansion}
The JT expansion, which is also known as the throttling process, occurs when a gas of high pressure section penetrates to a low pressure section through a porous plug and is a fundamentally irreversible process. In this process, the enthalpy is kept constant and the gas undergoes an adiabatic expansion, so an isenthalpic process can be applied to calculate the temperature change. This temperature change in the throttling process is encoded in the JT coefficient. That is, the numerical value of the slope of an isenthalpic curve on a $T\!-\!P$ diagram, at any point, is called the JT coefficient and is denoted by
\be\label{jtc1} \mu =\left(\frac{\prt T}{\prt P}\right)_{H}\,.\ee
The locus of all points for zero JT coefficient is known as the inversion curve. An important feature of JT coefficient is that the region in $T\!-\!P$ diagram where $\mu>0$, it is called the cooling region and where $\mu<0$, is called heating region. 
From the first law of black hole thermodynamics, the change in the mass is given by
\be\label{fl} dM=TdS+\Phi dQ+VdP\,,\ee
where $\Phi$ is the electric potential of the black hole at the horizon
\be\label{elpot} \Phi=\left(\frac{\prt M}{\prt Q}\right)_{S,P}=\frac{Q}{r}\,,\ee
for constant $S$ and $P$. Taking into account $dM=dQ=0$ in the throttling process, then
\be\label{HQ} T\left(\frac{\prt S}{\prt P}\right)_{M}+V=0\,,\ee
 where by using thermodynamic variation $dS\!=\!\left(\frac{\prt S}{\prt P}\right)_{T} dP+\left(\frac{\prt S}{\prt T}\right)_{P} dT$ and Maxwell relation $\left(\frac{\prt S}{\prt P}\right)_{T}\!=\!-\left(\frac{\prt V}{\prt T}\right)_{P}$ we obtain
 \be\label{jtc2} \mu=\frac{1}{C_{P}}\left[T\left(\frac{\prt V}{\prt T}\right)_{P}-V\right]\,,\ee
so if we substitute the relations (\ref{vol}), (\ref{temp2})-(\ref{ent}) in (\ref{jtc2}), and that $C_{P}=T\left(\frac{\prt S}{\prt T}\right)_{P,Q}$ it gives
 \be\label{JT} \mu=\frac{2}{3 k} \left(\frac{r^2}{\eta  G_{p}+r^2}\right)^{3/2} \frac{16 \pi  P r^6+ (8 \pi  \eta  G_{p} P+4)r^4+3(\eta  G_{p}-2Q^2) r^2- 5 \eta  G_{p} Q^2}{ r \left(8 \pi  P r^4+r^2-Q^2\right)}\,.\ee

\begin{figure}[H]
\centering
  \begin{subfigure}{0.45\textwidth}
    \includegraphics[width=\textwidth]{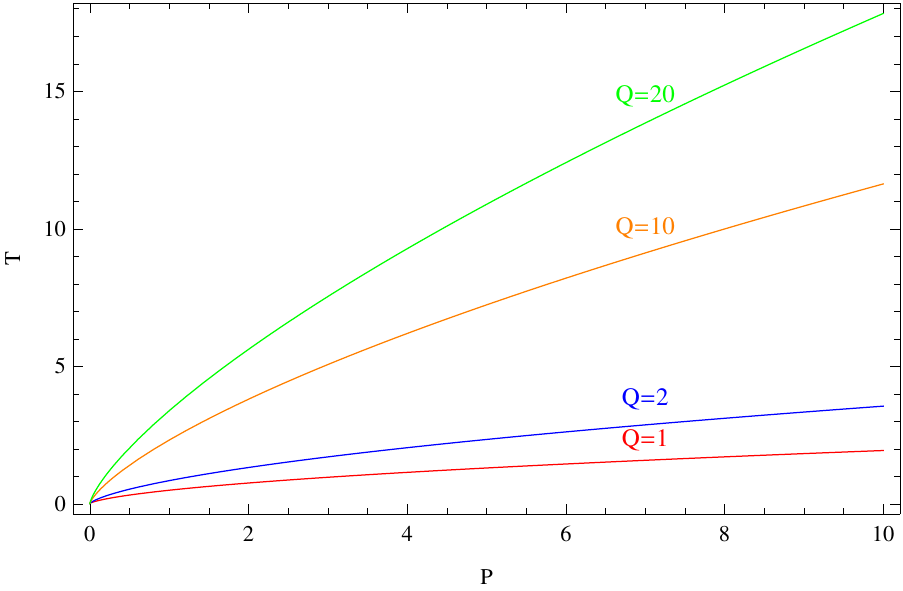}
    \caption{Large pressures}
    \label{fig1}
  \end{subfigure}
  \hspace{5mm}
  \begin{subfigure}{0.45\textwidth}
    \includegraphics[width=\textwidth]{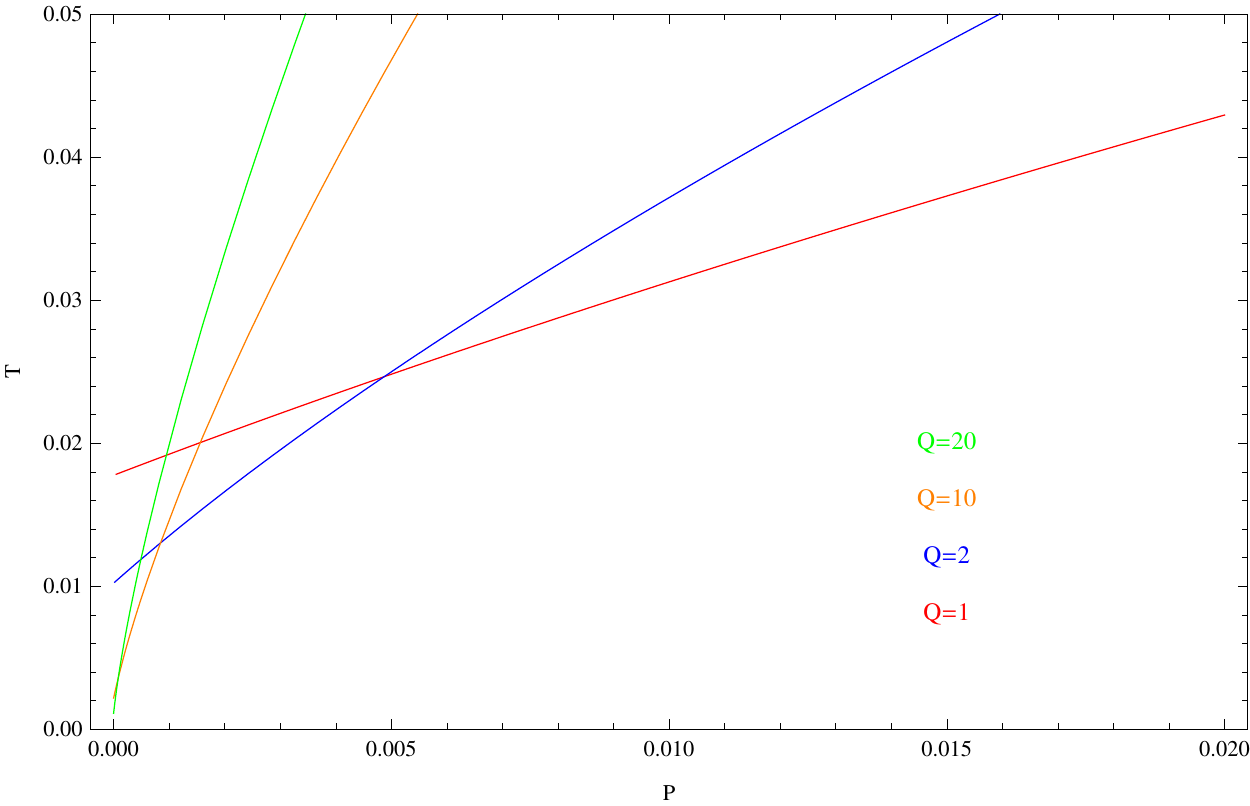}
    \caption{Low pressures}
    \label{fig2}
  \end{subfigure}
  \caption{\small{The inversion curves for charged AdS black hole in RG with $G_{p}=\eta=k=1$ but different charges. }}
  \label{f1}
\end{figure}

 The inversion curves $T_{i}$ versus $P_{i}$ are obtained by solving $\mu=0$, as a result of that we compute $r_{+}$ in terms of $P_{i}$ and then substitute the largest root in (\ref{temp2}). We have plotted the inversion curves for different values of black hole charge in Fig. (\ref{fig1}). As illustrated in figure, the graphs are monotonically increasing, so there is no maximum inversion temperature for charged AdS black holes in RG. This behavior is essentially different from the real gases in thermodynamics such as van der Waals gas \cite{Okcu:2016tgt,Okcu:2017qgo,Zhao:2018kpz}. Each curve has a minimum value of inversion temperature $T_{i}^{min}$, which corresponds to zero inversion pressure, $P_{i}=0$. Since $T_{i}(P)$ is a monotonically increasing function, there is only a minimum inversion temperature and the cooling and heating regions lie above and below the inversion curve, respectively.
Though the inversion curves have the same behavior for the large pressures, but as depicted in Fig. (\ref{fig2}) in the low pressure limit they have different behavior. That is by increasing the charges they fall to low temperatures.

\begin{figure}[h]
\centering
  \begin{subfigure}{0.43\textwidth}
    \includegraphics[width=\textwidth]{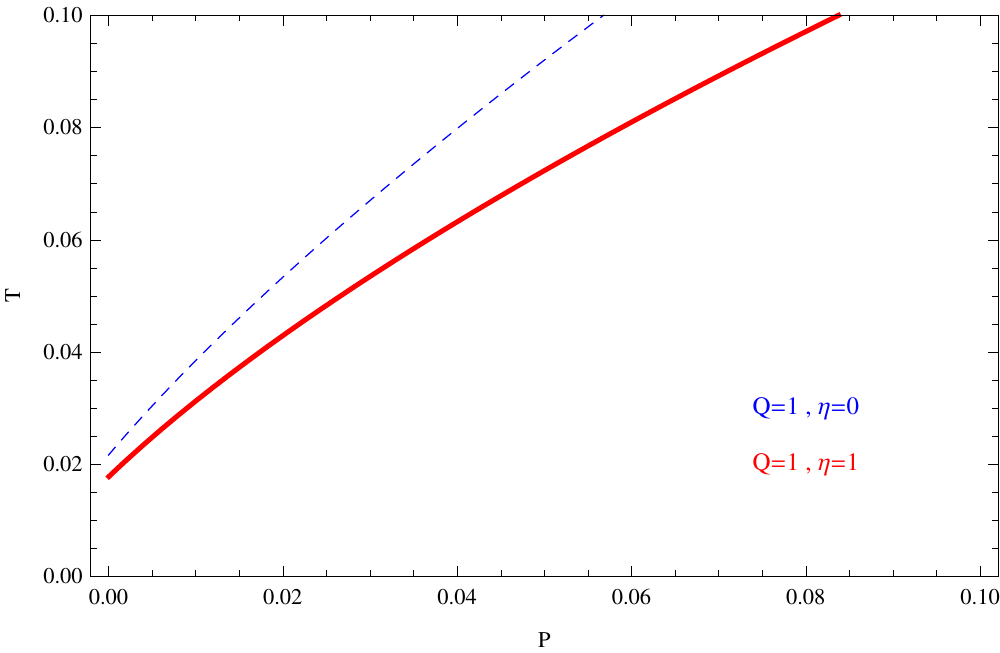}
    \caption{}
    \label{fig3}
  \end{subfigure}
  \hspace{5mm}
  \begin{subfigure}{0.43\textwidth}
    \includegraphics[width=\textwidth]{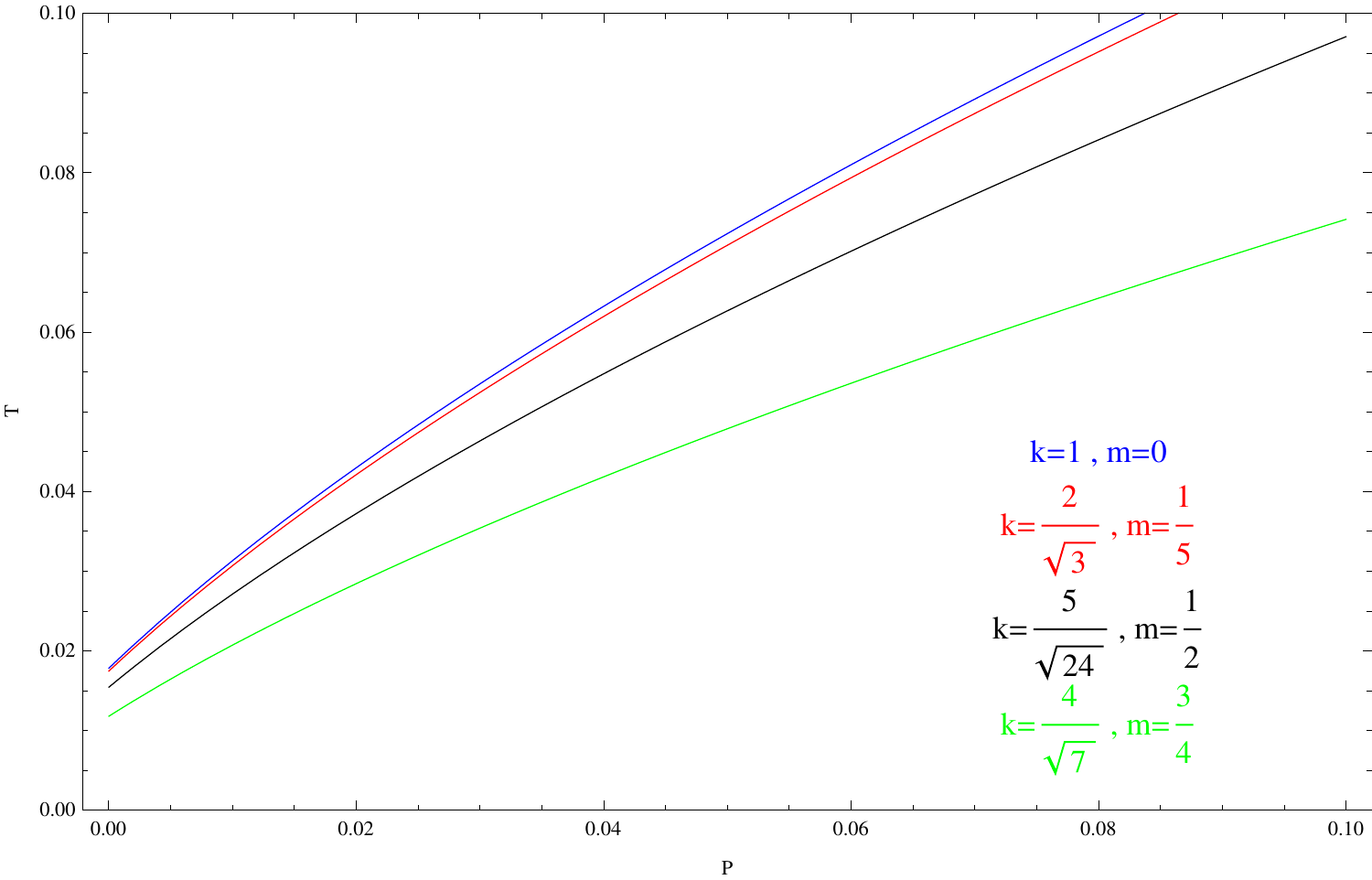}
    \caption{}
    \label{fig4}
  \end{subfigure}
  \caption{\small{(a) The inversion curve for RG with solid red line ($\eta=1$) has lower slope than the GR with dashed blue line ($\eta=0$). (b) The inversion curve of different masses for $Q=1$, $G_{p}=1$, and $\eta=1$.  }}
  \label{f2}
\end{figure}

\begin{figure}[H]
\centering
  \begin{subfigure}{0.45\textwidth}
    \includegraphics[width=\textwidth]{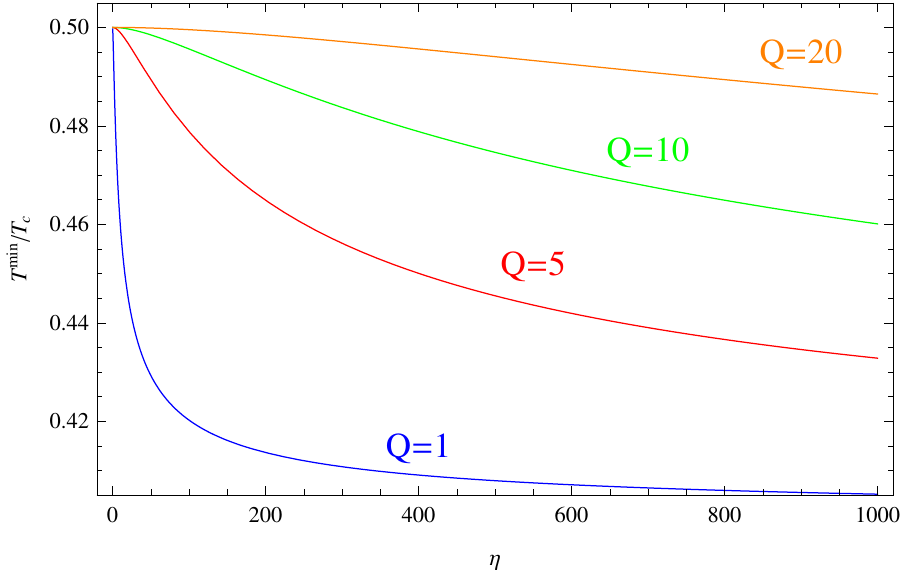}
    \caption{different charges}
    \label{fig5}
  \end{subfigure}
  \hspace{5mm}
  \begin{subfigure}{0.45\textwidth}
    \includegraphics[width=\textwidth]{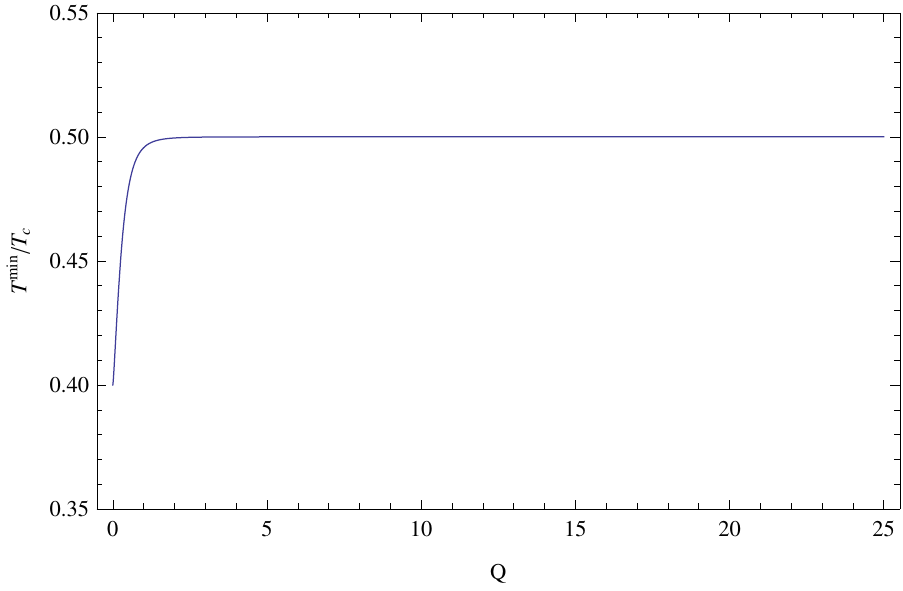}
    \caption{ $\eta=1$}
    \label{fig6}
  \end{subfigure}
  \caption{\small{The ratio of minimum inversion temperature to critical temperature. }}
  \label{f3}
\end{figure}

From the relation (\ref{JT}) if we use $P_{i}=0$ in $\mu=0$ equation and then solve it for $r_{+}$, we obtain four values for it. By substituting the largest one in (\ref{temp2}) we obtain the minimum inversion temperature
\be\label{tmin1}T_{i}^{min}\!=\!\!\frac{\left(18 Q^2+11 \eta  G_{p}-3 \sqrt{9 \eta ^2 G_{p}^2+44 \eta  G_{p} Q^2+36 Q^4}\right) \sqrt{\sqrt{9 \eta ^2 G_{p}^2+44 \eta  G_{p} Q^2+36 Q^4}+5 \eta  G_{p}+6 Q^2}}{40 \sqrt{2} \pi  \eta ^2 G_{p}^2 k},\ee
where if we expand it for small $\eta$
\be\label{tmin2}T_{i}^{min}=\frac{1}{6\sqrt{6} k \pi Q}-\frac{G_{p} \eta}{24\sqrt{6}k\pi Q^3}+\mathcal{O}(\eta^2),\ee
the leading term for $k=1$ gives the minimum inversion temperature in Ref. \cite{Okcu:2016tgt}.
We have compared the inversion curves for the charged AdS black holes in rainbow gravity with the one in general relativity  in Fig. (\ref{fig3}) and it shows that the slope of the RG's plot is less than the GR case. We have also plotted the inversion temperature curves for different values of the test particle's mass in Fig. (\ref{fig4}).
To better understand the behavior of this thermodynamic system we calculate the critical points from inflection point of $P=P(r_{+})$ \cite{Kubiznak:2012wp,Li:2018gwf}, i.e.,
 \be \frac{\prt P}{\prt r_{+}}=0\,,\qquad \frac{\prt^2 P}{\prt r_{+}^2}=0\,.\ee
 These equations lead to the critical values 
 \bea\label{critr}
 r_{c}&\!\!\!\!=\!\!\!\!&\sqrt{2} \left[Q^2+\left(\eta  G_{p} Q+Q^3\right)^{2/3}+\frac{\left(\left(\eta  G_{p} Q+Q^3\right)^2\right)^{2/3}}{\eta  G_{p}+Q^2}\right]^{\frac12}\,,\\
 \label{critt} T_{c}&\!\!\!\!=\!\!\!\!&\frac{1}{2\pi k \,r_{c}^2}\frac{r_{c}^2-2Q^2}{r_{c}^2+2\eta G_{p}} \sqrt{r_{c}^2+\eta G_{p}}\,,\eea
 where in the limit $\eta\rightarrow 0$ they give the critical values in \cite{Okcu:2016tgt}.
It has been shown in \cite{Okcu:2016tgt} that the ratio of minimum inversion temperature to the critical temperature for the charged AdS black holes in GR is equal one-half, but here we show in the case of RG this value will correct. As shown in Fig.(\ref{fig5}) this ratio is descending by growing $\eta$ while the rate of changes decreases by increasing the charge of black hole. A similar behavior occurs in Lovelock \cite{Mo:2018qkt} or in higher dimensional charged AdS black holes \cite{Mo:2018rgq} when we increase the Lovelock parameter or the dimensions. The Fig. (\ref{fig6}) describes that for a particular value of $\eta=1$ the ratio will deviate significantly for small $Q$ but it does not change for larger values. This also can be seen by expanding the ratio for small and large limit of the charge. While the leading term is 2/5 for small Q
  \be \frac{T_{i}^{min}}{T_{c}}=\frac25+\frac{3Q^{4/3}}{5}-\frac{13Q^2}{15}+\mathcal O \left(Q^{7/3}\right),\ee
 it is 1/2 for large Q, i.e.,
   \be \frac{T_{i}^{min}}{T_{c}}=\frac12-\frac{1}{108 Q^4}+\mathcal O\left(Q^{-6}\right).\ee

\begin{figure}[H]
\centering
  \begin{subfigure}{0.4\textwidth}
    \includegraphics[width=\textwidth]{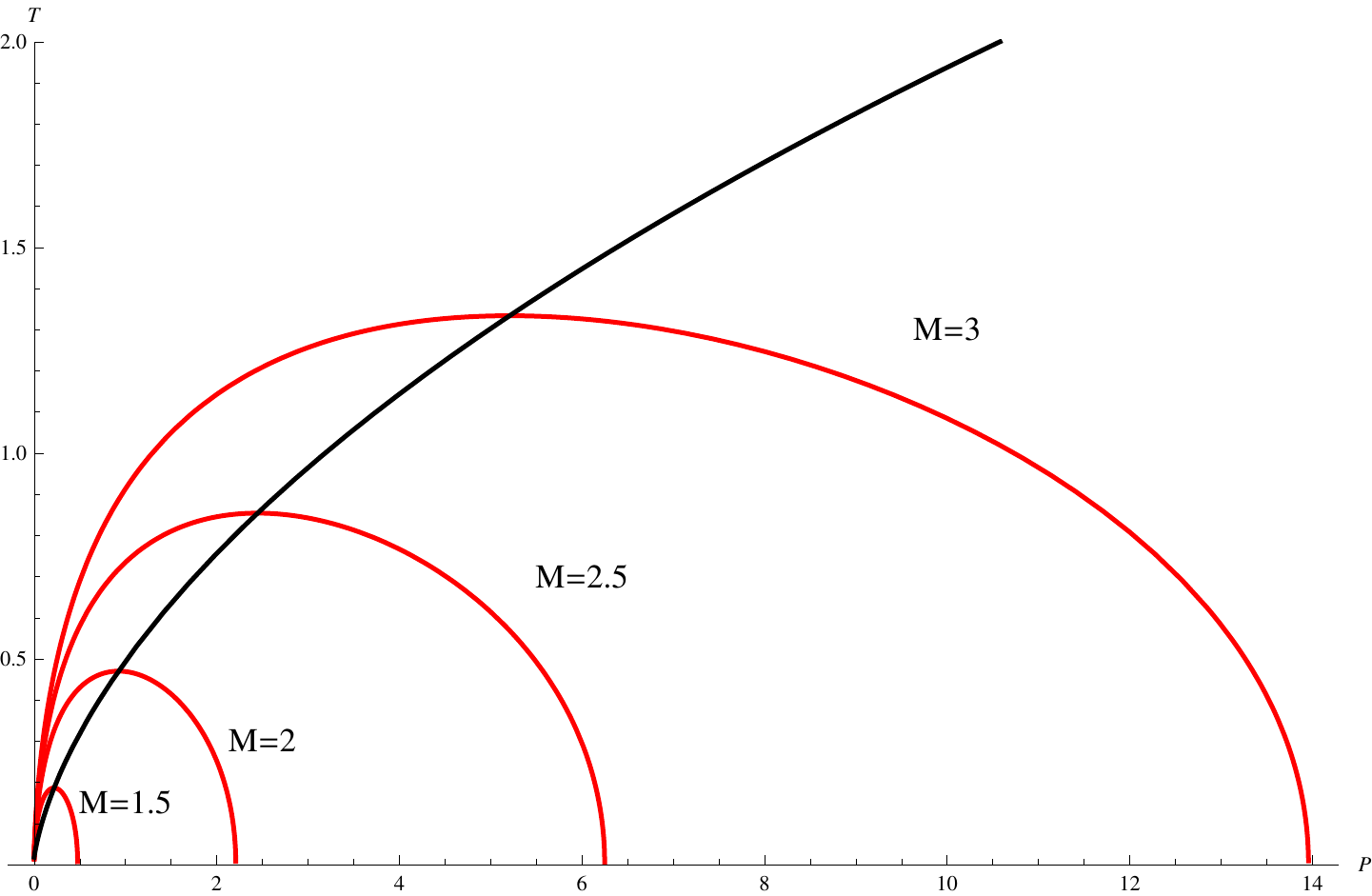}
    \caption{\small{$Q=1$}}
    \label{fig7}
  \end{subfigure}
  \hspace{5mm}
  \begin{subfigure}{0.4\textwidth}
    \includegraphics[width=\textwidth]{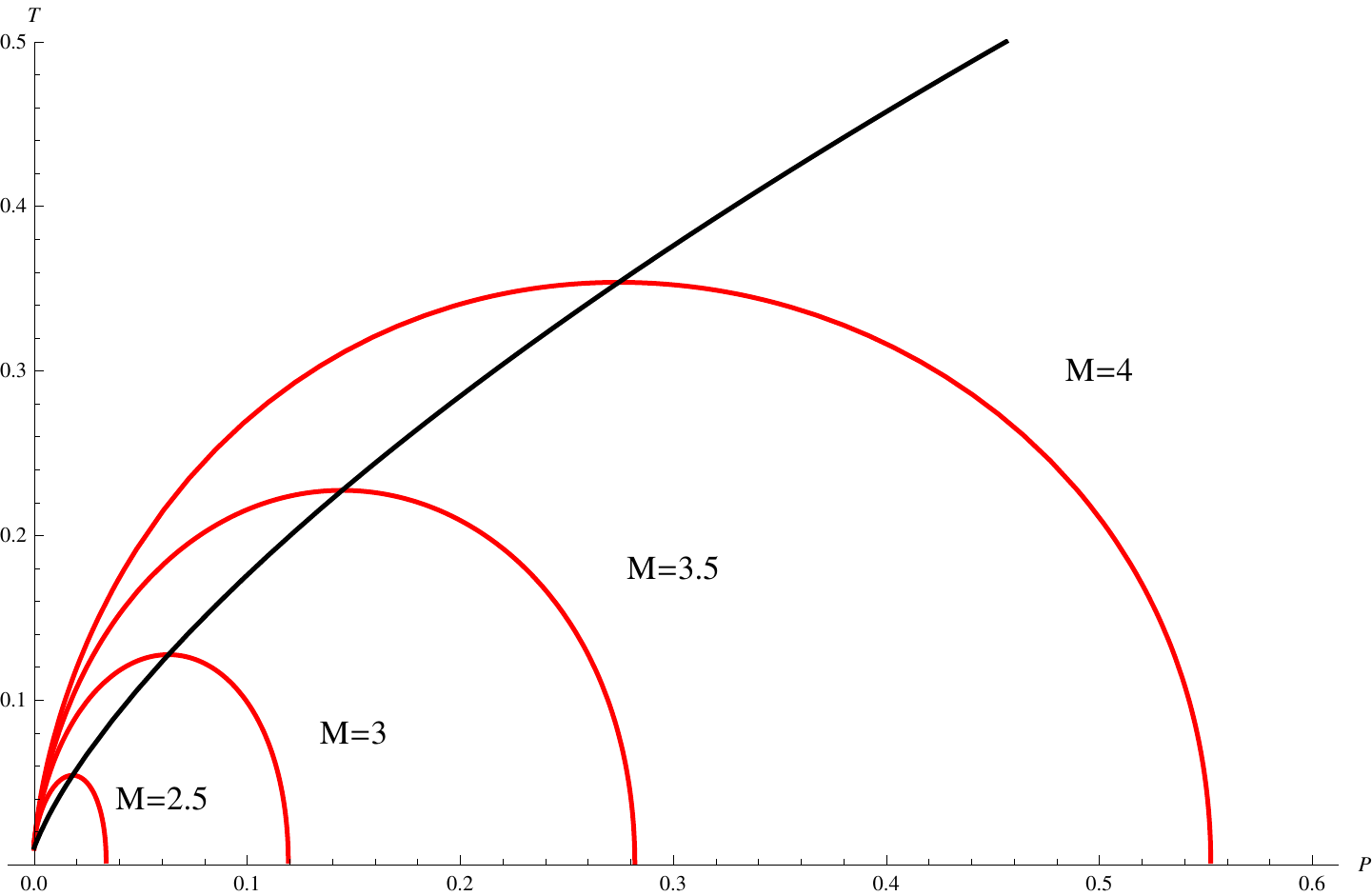}
    \caption{\small{$Q=2$}}
    \label{fig8}
  \end{subfigure}
   \begin{subfigure}{0.4\textwidth}
    \includegraphics[width=\textwidth]{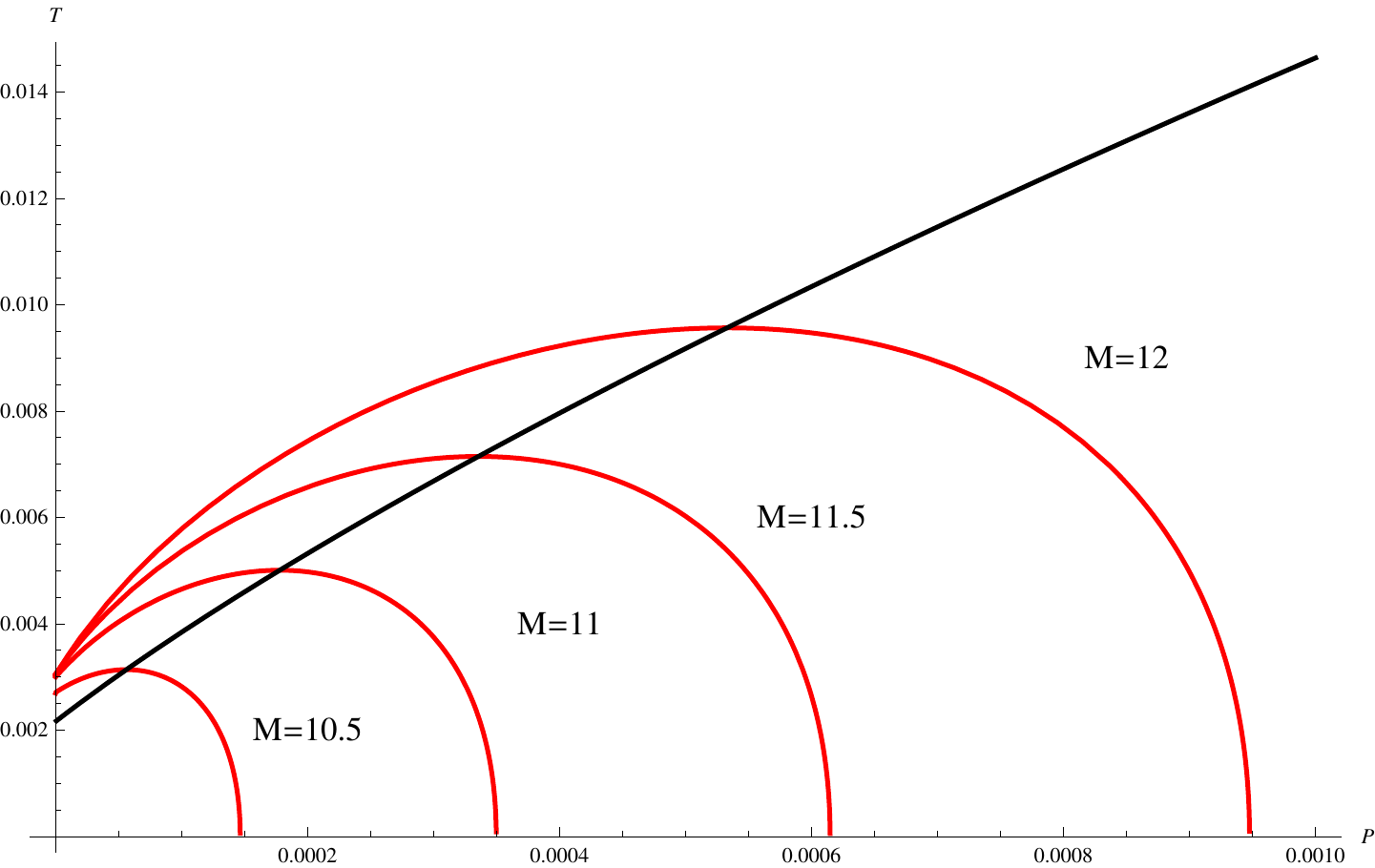}
    \caption{\small{$Q=10$}}
    \label{fig9}
  \end{subfigure}
  \hspace{5mm}
    \begin{subfigure}{0.4\textwidth}
    \includegraphics[width=\textwidth]{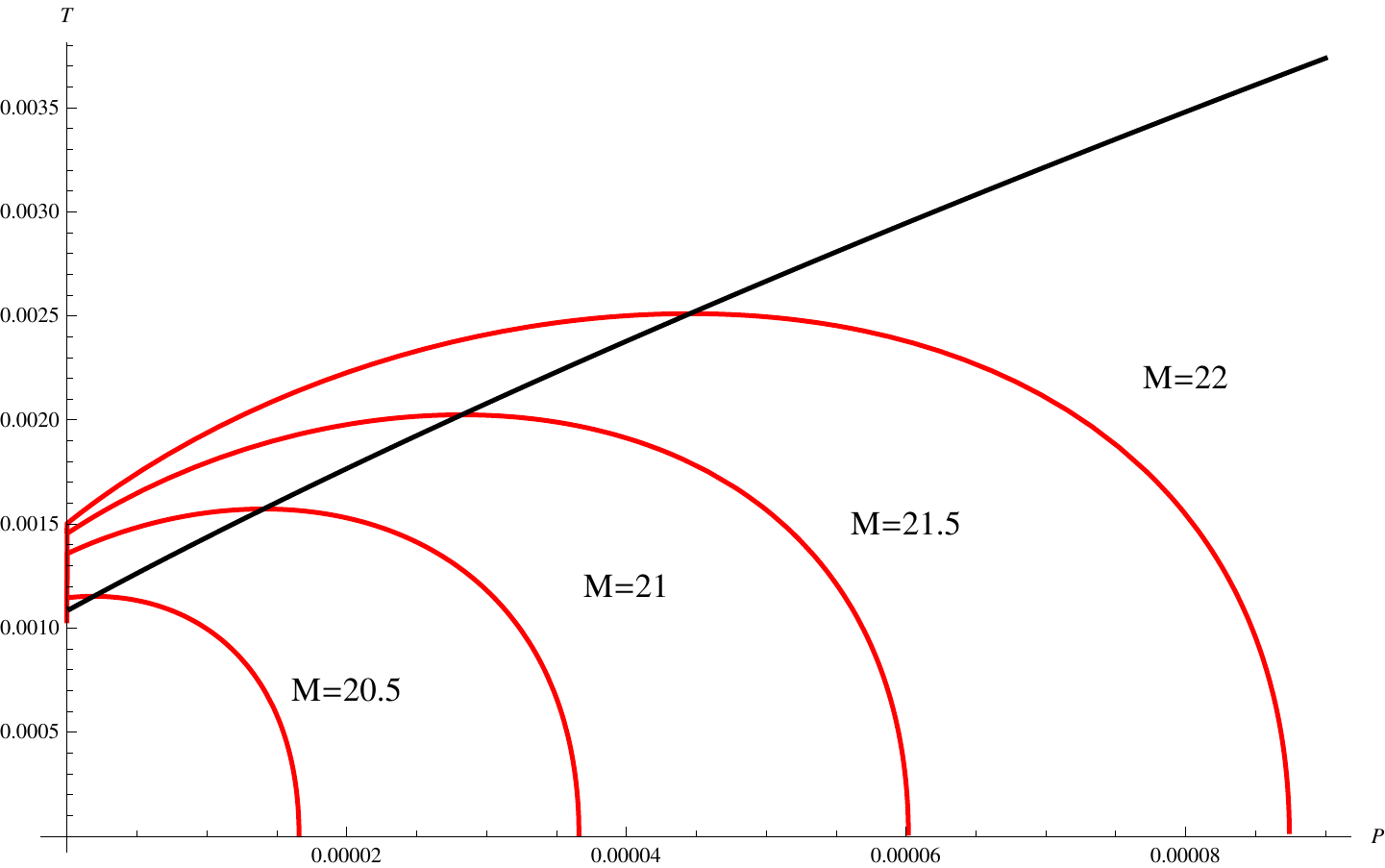}
    \caption{\small{$Q=20$}}
    \label{fig10}
  \end{subfigure}
  \caption{\small{The isenthalpic (red) and inversion (black) curves for $\eta=1$, $G_{p}=1$, and $k=1$. }}
  \label{f4}
\end{figure}

Since the JT expansion is an isenthalpic process, it is also of interest to consider the isenthalpic (constant mass) curves. We can obtain the isenthalpic curves in the $T-P$ plane by inserting the value of the event horizon in the equation of state (\ref{press}). The isenthalpic curves for different values of black hole charge have plotted in Figs.(\ref{f4}). The inversion temperature curve intersects them at their maximum points in each diagram. As seen from the figures the inversion curve intersects the isenthalpic curves at lower pressures as the charge increases and vice versa for the mass.

\begin{figure}[ht]
\centering
  \begin{subfigure}{0.45\textwidth}
    \includegraphics[width=\textwidth]{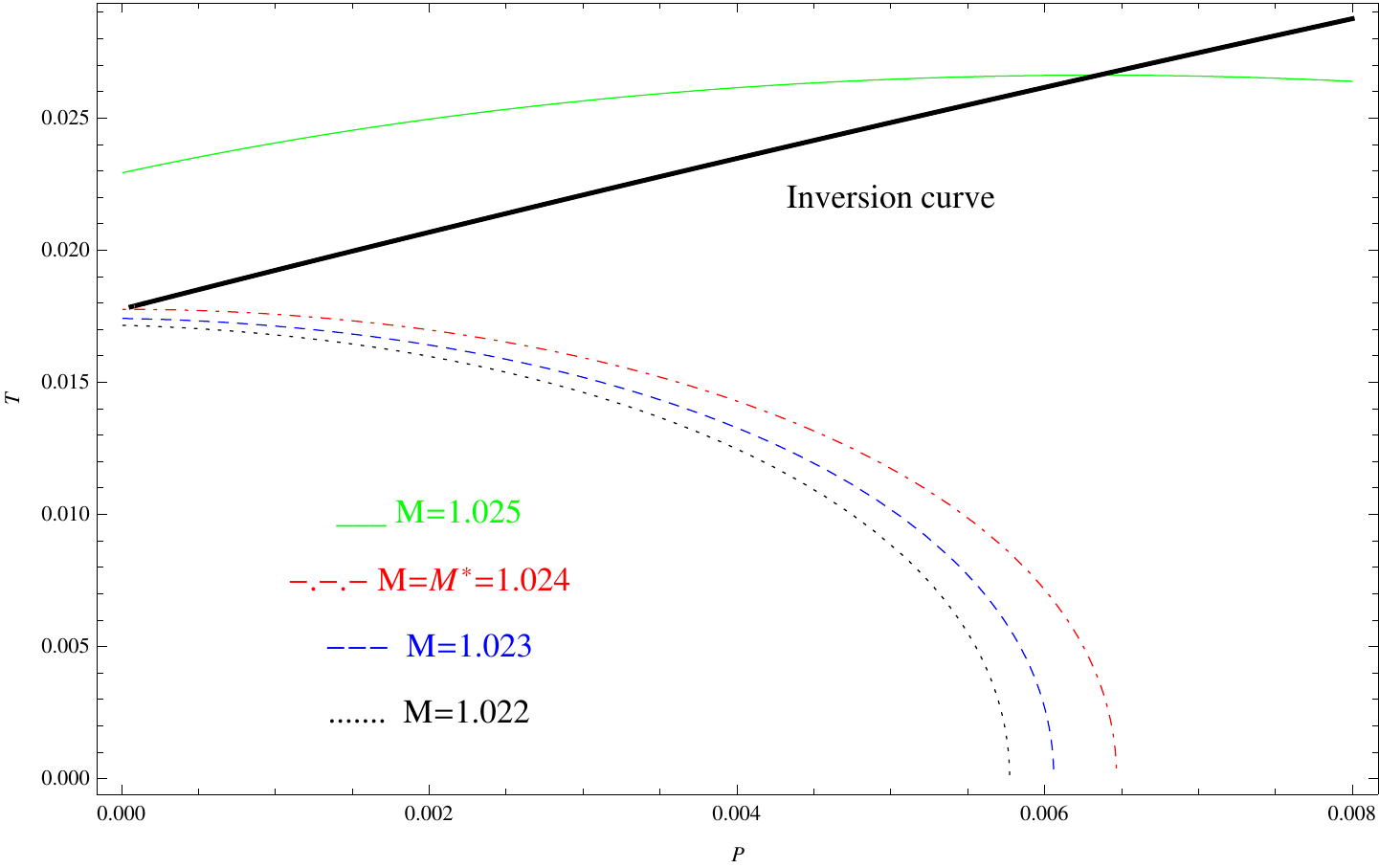}
    \caption{Intersection with inversion curve}
    \label{fig11}
  \end{subfigure}
  \hspace{5mm}
  \begin{subfigure}{0.45\textwidth}
    \includegraphics[width=\textwidth]{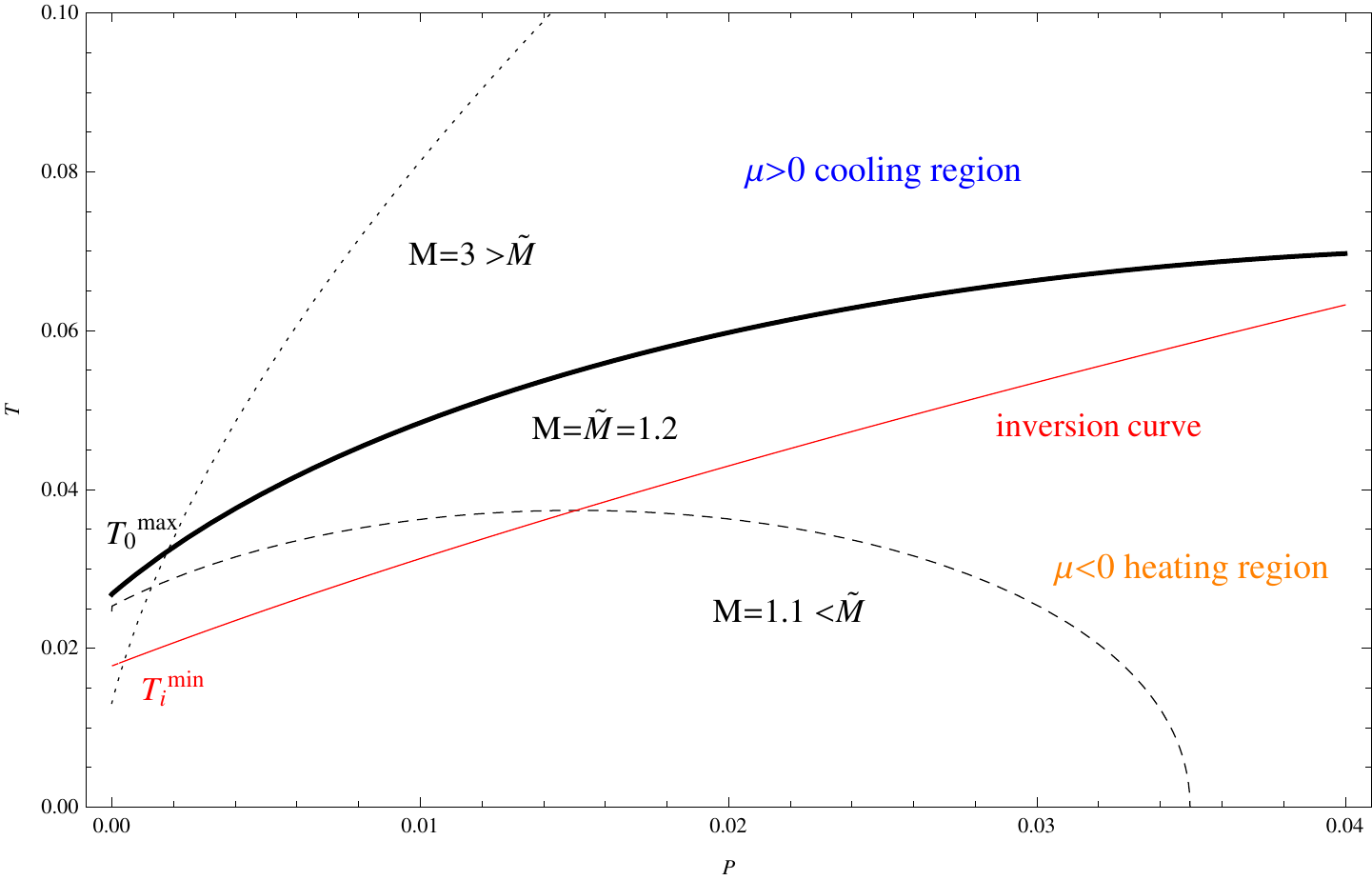}
    \caption{Intersection with each other}
    \label{fig12}
  \end{subfigure}
  \caption{\small{The behavior of isenthalpic curves for different values of black hole mass.}}
  \label{f5}
\end{figure}

We can also have the curves that do not intersect with the inversion curve or the ones that intersect with each other. In the former, the temperature for zero pressure ($P=0$) is denoted by $T_{\circ}$, so by equalizing it to the minimum inversion temperature $T_{i}^{min}$, since the inversion curve is monotonically increasing, we can find a particular value of black hole's mass which we name it $M=M^{*}$. This curve is illustrated in Fig.(\ref{fig11}) by a red dashed dotted graph. As seen in this figure, the isenthalpic curve with $T_{\circ}>T_{i}^{min}$ or $M>M^{*}$ first rises and then after intersecting at extreme point with inversion curve descends, as well as plots in Figs.(\ref{f4}). On the other hand, for the curves with $T_{\circ}<T_{i}^{min}$ or $M<M^{*}$ there is no inversion point in heating region. In the latter curves, in the zero pressure limit, there is also a maximum value for the temperature $T_{\circ}^{max}$ which is correspond to a particular mass $M=\tilde{M}$. We calculate these masses for the case $Q=1$ as $M^{*}=1.024$ and $\tilde{M}=1.208$. For any isenthalpic curve with a mass greater than $\tilde{M}$ the curve should intersect with the others in the $T-P$ plane, as shown in Fig.(\ref{fig12}) by the doted curve for $M>\tilde{M}$.
%%%%%%%%%%%%%%%%%%%%%%%%%%%%%%%%%%%%%%%%%%%%
\section{Conclusion}
In this letter we have mainly studied the Joule-Thomson (JT) expansion process in the rainbow gravity (RG) for charged AdS black holes. In this process, which describes the expansion of gas from a high pressure section to a low pressure section through a porous plug, we obtained an exact expression for the JT coefficient $\mu$ by an analytical recipe from the first law of black hole thermodynamics and investigated its behavior intuitively by plotting the inversion and isenthalpic curves. The mass of black hole is kept constant during this expansion as an isenthalpic process. The inversion curves ($\mu=0$) depicted in Figs.(\ref{fig1},\ref{fig2}) show that they have different behavior in low and large pressures and the slope of curves increases by growing the charge of black hole. We have also shown that the slope of inversion curves slows down in RG by increasing both of the rainbow parameter $\eta$ and the test particle's mass $m$ in Figs.(\ref{fig3},\ref{fig4}) in contrast to the effect of charge Q. One of the remarkable by-product studies in this paper is the ratio between the minimum inversion temperature and critical temperature where in the case of charged AdS black holes is equal one-half. Moreover, we have indicated in Figs.(\ref{fig5},\ref{fig6}) that this ratio will change from this value by considering $\eta$ and $Q$. 

One of the significant subjects in this work was the analyzing of the isenthalpic curves during the JT expansion. For each curve, the highest value of temperature occurs at maximum point where $\mu=0$ and is called the inversion point, discriminating the cooling from heating process. According to Figs. (\ref{f4}) the higher inversion temperature for JT expansion corresponds with larger enthalpy (mass) and for larger charges the inversion point appears in lower temperature or pressure. The difference between these curves with the ones in \cite{Okcu:2016tgt} shows how the rainbow parameter affect the exhausting/absorbing heat process to the reservoir in cooling/heating regions, respectively. We inferred from the isenthalpic curves in Fig.(\ref{fig11}) that there is an upper bound for the mass of black hole for which we could have an inversion point, that is, for $M<M^{*}$ there is no inversion point and the expansion is always in the regime of heating process, where $M^{*}$ is determined from $T_{i}^{min}=T_{\circ}(p=0)$. We have also shown that the curves with $M>\tilde{M}$, where $\tilde{M}$ comes from the extremum of $T_{\circ}(p=0)$, intersect with the other isenthalpic curves before crossing the inversion curve in the cooling region. 
%%%%%%%%%%%%%%%%%%%%%%%%%%%%%%%%%%%%%%%%%%%%%%%
%\appendix
%%%%%%%%%%%%%%%%%%%%%%%%%%%%%%%%%
%\section*{Acknowledgment}
%@@@@@@@@@@@@@@@@@@@@@@@@@@@@@@@@@@@@@@@@@@@@@@@@@@@@@@@@@@@@@@@@@@@@@@@@@@@@@@@@@@@@
%%%%%%%%%%%%%%%%%%%%%%%%%%%%%

\end{document}